\documentclass[preprint, aps,amsmath,amssymb,floats,12pt]{revtex4}
\usepackage{amsmath}
\usepackage{graphicx}% Include figure files
\usepackage{dcolumn}% Align table columns on decimal point
\usepackage{bm}% bold math
\usepackage[dvips]{color}
\begin{document}
%opening

\title{A Unified View of Transport Equations}

\author{J.A. Secrest$^1$, J.M. Conroy$^2$ and  H.G. Miller$^{2}$ }

\affiliation{$^1$Department of Physics and Astronomy, Georgia Southern University, Armstrong Campus, Savannah, Georgia, USA}

\affiliation{$^2$Department of Physics, State University of New York at Fredonia, Fredonia, New York,
USA}

\date{\today}

\begin{abstract}
Distribution functions of many static transport equations are found using the Maximum 
Entropy Principle. The equations of constraint which contain the relevant 
dynamical information are simply the low-lying moments of the distributions. 
Systems subject to conservative forces have also been considered.
\end{abstract}
\maketitle
\section{Introduction}
Transport phenomena is the description of how physical quantities such as the 
number of particles or the  energy flow through a medium.  In general these 
phenomena 
are described by various differential equations, for example, the Boltzmann 
equation, diffusion equation, and the advection equation \cite{tp1,tp2,tp3}. 
In the simplest case the solutions of many of these equations, i.e. the 
distribution 
functions,  $\rho(x)$, have 
one thing in common in that they are simple Gaussian distributions which do not change 
as a function of time.
At any given point in time or in the static case, these distributions are 
uniquely determined by knowledge of at most two moments. This strongly suggests 
that the relevant dynamical information  to be included is contained in the 
moments. Rather than model these 
phenomena by constructing a 
differential equation in each case one can use the Maximum Entropy 
Principle(MEP)]\cite{shannon1,jaynes}  
where the dynamics are contained in the equations of constraint.

In the static case the MEP requires that the information entropy, $S(\rho)$, 
satisfies the equation
\begin{equation}
\label{eqn:max_S}
\delta \rho S(\rho) = 0
\end{equation}
subject to the relevant equations of constraint
\begin{equation}
 Tr(\rho(x)\ O_i(x)) =r_i.
 \label{eoc4}
\end{equation}
Classically
\begin{equation}
 S(\rho) = -\int_{-\infty}^{\infty}\rho(x) 
\text{ln} \frac{\rho(x)}{\rho_{0}(x)} dx
\end{equation}
and $\rho_0(x)$ is an invariant measure.
The solution to the above equations is given by
\begin{equation}
 \rho(x)=\rho_0(x) e^{\sum_{i=0}\lambda_i O_i(x)}
  \label{MEPSoln}
\end{equation}
and the Lagrange multipliers, $\lambda_i$, are determined from the equations 
of constraint.

In the simplest cases, as we shall show, the relevant choice for $O_i(x)$ is 
given by the lowest moments of $x$
\begin{equation}
 O_i(x) = x^i
\end{equation}
for $i=0,1,2$ as this will ultimately yield Gaussian distributions since the 
$i=1$ 
component will not contribute.

Furthermore if the system is acted on by an external conservative force derivable from a potential,
$K(x)=-\frac{\partial u(x)}{\partial x}$, such that
\begin{equation}
 \frac{dx}{dt}=K(x),
\end{equation}
the accompanying Liouville equation is of the form\cite{PP95}
\begin{equation}
 J(x)=\rho(x)K(x)
\end{equation}
where the current $J(x)$ satisfies the continuity equation
\begin{equation}
 \frac{\partial \rho(x)}{\partial t}+ \nabla \cdot J(x)=0.
 \end{equation}
 
 In the static case the current $J(x)$ is constant. Since $\rho(x)\rightarrow 0$ as $x 
\rightarrow \pm \infty$
 the constant must be $0$. Hence
 \begin{equation}
  \rho(x)\frac{\partial u(x)}{\partial x}=0.
 \end{equation}
Integrating by parts where the initial potential $u(x_0)=0$ yields
\begin{equation}
 \rho(x) u(x)=\int^x_{x_0}\frac{\partial \rho(x)}{\partial x} u(x) dx
\end{equation}
or 
\begin{equation}
Tr[ \rho(x) u(x)]=Tr\Big[\int^x_{x_0}\frac{\partial \rho(x)}{\partial x} u(x) dx\Big],
\end{equation}
which is the proper form for an equation of constraint (see 
Eq.(\ref{eoc4})).

Hence the general form of the distribution functions associated with many of 
the well known transport equations is given by
\begin{equation}
 \rho(x)=e^{\sum_{i=0}\lambda_i x^i + \lambda_3 u(x)}    
  \label{MEPSoln_mod}
\end{equation}
with $i=0,1,2$. The values of the Lagrange multipliers are determined by the 
relevant equations of constraint which  contain the information about the 
particular dynamics involved.

\section{Examples}
In this section the use of the MEP for solving transport phenomena equations is 
demonstrated through a number of examples. The analytic solution is known in each of the following examples.  In practice, the necessary constraints would be 
determined experimentally and applied to the MEP general solution.  For 
example, 
the second spatial time-dependent moment of the diffusion equation studied 
below 
can be determined by measuring the diffusion constant and invoking the Einstein 
relation as discussed below.  Another example would be to determine the velocity moments by measuring the momentum distributions. In the following 
examples one spatial dimension is considered, though the MEP can be extended to 
higher dimensions. 

\subsection{Advection Equation}
The advection equation is a hyperbolic partial differential equation of the form
\begin{equation}
\label{eqn:advection_eqn}
\frac{\partial \rho(x,t)}{\partial t} = -v \frac{\partial \rho(x,t)}{\partial x}
\end{equation}
that describes how a scalar field density $f$ is swept along (advected) by a 
bulk flow of constant speed $v$.  The position $-\infty < x < \infty$, the time 
$0 < t < \infty$, and the velocity field $v$ are nonzero. Examples of where the 
advection equation is used are modeling automobile traffic, blood flow through 
a 
capillary, and salinity propagation in the ocean. 
 
%The general solution to this equation is of the form
%\begin{equation}
%\rho(x,t)=\rho_{0}(x-vt)
%\end{equation}    
%where $\rho_{0}=\rho(x,0)=\rho_{0}(x)$ is the initial condition.
A particular solution to this equation is 
\begin{equation}
\label{eqn:adv_soln}
\rho(x,t) = e^{-(x-vt)^2}.
\end{equation}
%with initial condition $\rho_{0}=e^{-x^2}$.

The moments calculated using the actual solution (which would be experimentally 
determined) are:
\begin{eqnarray}
\label{eqn:m_0}
r_{0} &=& \int_{-\infty}^{\infty} e^{(x-vt)^{2}} dx = \sqrt{\pi}, \\
\label{eqn:m_1}
r_{1} &=& \int_{-\infty}^{\infty} xe^{(x-vt)^{2}} dx = \sqrt{\pi}vt, \\
\text{and}\nonumber \\
\label{eqn:m_2}
 r_{2} &=& \int_{-\infty}^{\infty} x^{2}e^{(x-vt)^{2}} dx = \frac{1}{2} 
\sqrt{\pi}[1+2v^2t^2].
\end{eqnarray}

The moments calculated from the generalized MEP solution seen in Eq. \ref{MEPSoln} are:
\begin{eqnarray}
r_{0} &=& \int_{-\infty}^{\infty}e^{-\lambda_{2}x^{2} - \lambda_{1}x - 
\lambda_{0}}dx = e^{-\lambda_{0}} 
\bigg[\frac{e^\frac{\lambda_{1}^{2}}{4\lambda_{2}} \sqrt{\pi}} 
{\sqrt{1+\lambda_{2}} } \bigg],\\
r_{1} &=& \int_{-\infty}^{\infty}xe^{-\lambda_{2}x^{2} - \lambda_{1}x - 
\lambda_{0}}dx = e^{-\lambda_{0}} 
\bigg[-\frac{\lambda_{1}e^\frac{\lambda_{1}^{2}}{4\lambda_{2}} \sqrt{\pi}} 
{2(1+\lambda_{2})^{3/2} } \bigg],\\
\text{and}\nonumber \\
r_{2} &=& \int_{-\infty}^{\infty}x^{2}e^{-\lambda_{2}x^{2} - \lambda_{1}x - 
\lambda_{0}}dx = e^{-\lambda_{0}}
\bigg[\frac{(2\lambda_{2}+\lambda_{1}e^\frac{\lambda_{1}^{2}}{4\lambda_{2}} 
\sqrt{\pi}} {4(1+\lambda_{2})^{5/2} } \bigg].
\end{eqnarray}
Equating the moments and solving for the Lagrange multipliers, 
$\lambda_{0}=(vt)^{2}$, $\lambda_{1}=-2vt$, and $\lambda_{2}=1$.  Substituting 
these results back into the MEP solution Eq.(\ref{MEPSoln}) results in the 
solution given by Eq.(\ref{eqn:adv_soln}).
	
\subsection{Diffusion Equation}
This choice of the diffusion equation is a linear partial differential equation 
with a constant diffusion constant $D$ with the form
\begin{equation}
\label{eqn:diffusion_eqn}
\frac{\partial \rho(x,t)}{\partial t} = D \frac{\partial^{2} 
\rho(x,t)}{\partial 
x^{2}}
\end{equation}
that describes how a scalar field density $\rho$ spreads out as a function of 
position $x$ and time $t$. This equation describes the collective motion of 
random particles.  The same equation also describes heat flow,  financial 
markets and free particles in non-relativistic quantum mechanics. 
After some time $t$, the particular solution (where the diffusion constant has been set equal to one) is
\begin{equation}
\label{eqn:diff_soln_2}
\rho(x,t) = 
\frac{\sqrt{2\pi\sigma}}{\sqrt{\pi(2\sigma+4t})}e^{-\frac{x^2}{2\sigma+4t}}
\end{equation}
with the initial condition that
\begin{equation}
\rho_{0}(x,0) = e^{\frac{x^2}{2\sigma}}.
\end{equation}

Notice that the MEP solution must require the prefactor due to the initial 
condition.  It will also be discovered that the linear term is zero.  In this 
case,
\begin{equation}
\label{eqn:mod_max_ent}
\rho=\rho_{0}e^{-\lambda_{2}x^2-\lambda_{1}x-\lambda_{0}}=e^{-(\frac{1}{2\sigma}
+\lambda_{2})x^{2}-\lambda_{0}}.
\end{equation}
The moments calculated from the analytic solution (which would be 
experimentally 
measured) are:
\begin{eqnarray}
r_{0} &=& 
\int_{-\infty}^{\infty}\frac{\sqrt{2\pi\sigma}}{\sqrt{\pi(2\sigma+4t})}e^{-\frac
{x^2}{2\sigma+4t}} = \sqrt{2\pi\sigma}\\
r_{1} &=& \int_{-\infty}^{\infty} 
x\frac{\sqrt{2\pi\sigma}}{\sqrt{\pi(2\sigma+4t})}e^{-\frac{x^2}{2\sigma+4t}} = 
0\\
\text{and}\nonumber \\
r_{2} &=& \int_{-\infty}^{\infty} 
x^{2}\frac{\sqrt{2\pi\sigma}}{\sqrt{\pi(2\sigma+4t})}e^{-\frac{x^2}{2\sigma+4t}} 
= \sqrt{2\pi\sigma}(2t+\sigma).
\end{eqnarray}

The moments calculated from the MEP solution Eq. (\ref{MEPSoln}) are:
\begin{eqnarray}
r_{0} &=& \int_{-\infty}^{\infty} 
e^{-(\frac{1}{2\sigma}+\lambda_{2})x^{2}-\lambda_{0}} = 
e^{-\lambda_{0}}\frac{\sqrt{2\pi}}{2\lambda_{2}+\frac{1}{\sigma}}\\
r_{1} &=& \int_{-\infty}^{\infty} x 
e^{-(\frac{1}{2\sigma}+\lambda_{2})x^{2}-\lambda_{0}} =0 \\
\text{and}\nonumber \\
r_{2} &=& \int_{-\infty}^{\infty} x^{2} 
e^{-(\frac{1}{2\sigma}+\lambda_{2})x^{2}-\lambda_{0}} = 
e^{-\lambda_{0}}\frac{\sqrt{2\pi}}{(2\lambda_{2}+\frac{1}{\sigma})^{\frac{3}{2}}
}.\\
\end{eqnarray}

Note that the first moment is zero from both calculations, indicating that 
there 
is no linear term in the exponent above.  Equating the moments and solving for 
the Lagrange multipliers, it is found that 
$e^{-\lambda_{0}}=\sqrt{\frac{\sigma}{2t+\sigma}}$ and $\lambda_{2} = 
-\frac{2t}{2\sigma(2t+\sigma)}$.  Substituting back the solutions for the 
Lagrange multipliers into Eq.(\ref{eqn:mod_max_ent}) results in the 
solution 
given by Eq.(\ref{eqn:diff_soln_2}).  In this example, the second spatial 
moment 
can be determined from the Einstein-Smoluchowski relation for the diffusion 
constant \cite{Einstein, Smoluchowski}.

\subsection{Fokker-Planck Equation with Logarithmic Potential}
The Fokker-Planck Equation is given by 
\begin{eqnarray}
\frac{\partial f(x,t)}{\partial t}=D\Big[\frac{\partial}{\partial 
x}K(x)-\frac{\partial^{2}}{\partial x^{2}}\Big]\rho(x,t) \nonumber
\end{eqnarray}
where $\rho(x,t)$ is the probability distribution function, $K(x)$ is related 
to 
the external force acting on the particles, and $D$ is the diffusion constant.  
The forcing function is related to a potential function $u(x)$ measured in 
units 
of Boltzmann temperature $k_{B}T=1$. One choice for the potential which can be solved for 
analytically 
is that of a logarithmic potential:
\begin{eqnarray}
u(x)= -U_{0}\text{ln}(|x|/a) 
\end{eqnarray}
where once again the diffusion constant has been set equal to one.
This yields the steady state solution \cite{FPE_log}
\begin{eqnarray}
\label{eq:steady_state_FPE}
\rho(x) = \frac{U_{0}-1}{2a}\Big(\frac{|x|}{a}\Big)^{-U_{0}}.
\end{eqnarray}
The zeroth moment and the expectation value of the potential energy were 
calculated to be
\begin{eqnarray}
r_{0} &=& 2\int_{a}^{\infty} 
\frac{1}{\frac{2a}{U_{0}-1}}e^{-U_{0}\text{ln}(\frac{x}{a})}=1 \\
\overline{u(x)} &=& 
2\int_{a}^{\infty}U_{0}\text{ln}\Big(\frac{x}{a}\Big)e^{-U_{0}\text{ln}(\frac{x
}{a})} \nonumber \\
&=&\frac{U_{0}}{U_{0}-1}.
\end{eqnarray} 
Determining the zeroth moment and the expectation value of the potential energy 
using the MEP solution Eq.(\ref{MEPSoln_mod}), it is found that
\begin{eqnarray}
r_{0} &=& 2 \int_{a}^{\infty} 
e^{\lambda_{0}+\lambda_{3}U_{0}\text{ln}(\frac{x}{a})} \nonumber \\
&=&2C\int_{a}^{\infty}e^{\lambda_{3}U_{0}\text{ln}(\frac{x}{a})} \nonumber \\ 
&=& -\frac{2aC}{1+\lambda_{3}U_{0}} \\
\overline{u(x)} &=& 2 \int_{a}^{\infty} U_{0}\text{ln}\Big(\frac{x}{a}\Big) 
e^{\lambda_{0}+\lambda_{3}U_{0}\text{ln}(\frac{x}{a})} \nonumber \\
 &=& 2C \int_{a}^{\infty} 
U_{0}\text{ln}\Big(\frac{x}{a}\Big)e^{\lambda_{3}U_{0}\text{ln}\frac{x}{a})} 
\nonumber \\
 &=& \frac{2aCU_{0}}{(1+\lambda_{3}U_{0})^{2}}
\end{eqnarray}
where the normalization constant has been rewritten as $e^{\lambda_{0}}=C$.  
Solving for the MEP solution using the undetermined multipliers $\lambda_{0}$ and $\lambda_{3}$ yield 
the 
steady state solution Eq.(\ref{eq:steady_state_FPE}). It is clear that no other moments contribute to constraining of the probability distribution function.

\section{Conclusions}
 A unified view of  
transport equations is possible within the framework of MEP.  The MEP solution 
that maximizes the information entropy with respect 
to the constraints is seen in Eq.(\ref{MEPSoln_mod}).  These constraints are typically 
moments of the distribution function but they may also be  constrained by other 
physical parameters such as a conservative potential. In the case of  diffusive phenomena 
such as Brownian motion, one  can determine the temporal evolution of the the 
second spatial 
moment from  measurements of the diffusivity, $<x^{2}>=r_{2}=2Dt$.  

A general form of the solution of distribution function for 
various transport phenomena has been found using the MEP.  Various  low-lying
moments
have been  used as the equations of constraint. 
This was applied to the advection equation and the diffusion equation. For 
distributions that are constant in time  
the time-dependent solutions along with any initial conditions that were 
specified
are used as these constraints.   It was shown that the potential 
energy of a conservative force could be used as a constraint on the 
distribution function.  This was illustrated in determining the steady-state 
solution to the Fokker-Planck equation with a logarthimic potential.  It should 
be noted that this yields an interesting solution that is ultimately a 
power-law 
solution instead of the more frequently encountered exponential-type solutions 
found with the MEP (though it should be noted that all types of distributions 
have been determined with MEP). Finally, using this technique to constrain the 
distribution from the potential opens up a number of different solutions to be 
found from transport equations that have conservative source terms. Lastly the 
MEP can be extended to obtain time dependent solutions \cite{MMPS99}.

\end{document}